\documentclass[a4paper,conference]{IEEEtran}
\IEEEoverridecommandlockouts
\usepackage{cite}
\usepackage{amsmath,amssymb,amsfonts}
\usepackage{algorithmic}
\usepackage{graphicx}
\usepackage{textcomp}
\usepackage{xcolor}
\def\BibTeX{{\rm B\kern-.05em{\sc i\kern-.025em b}\kern-.08em
    T\kern-.1667em\lower.7ex\hbox{E}\kern-.125emX}}

\usepackage{amssymb} 

\usepackage{xspace}

\newcommand{\keepcomment}{1} 

\usepackage[normalem]{ulem}

\ifnum\keepcomment=1
	\usepackage[colorinlistoftodos,textsize=scriptsize]{todonotes} 
    \newcommand{\stkout}[1]{\ifmmode\text{\sout{\ensuremath{#1}}}\else\sout{#1}\fi}
    
\else
	\newcommand{\todoi}[1]{\leavevmode\ignorespaces\unskip}
	
	\usepackage[disable]{todonotes} 
\fi

\newcommand{\todoi}[1]{\todo[inline]{#1}}
\usepackage{paralist} 

\begin{document}

\title{Parsimonious Edge Computing to Reduce Microservice Resource Usage}

\author{\IEEEauthorblockN{
Mathieu Simon\IEEEauthorrefmark{1},   
Alessandro Spallina\IEEEauthorrefmark{2},   
Loïc Dubocquet\IEEEauthorrefmark{1},    
Andrea Araldo\IEEEauthorrefmark{1}      
}                                     
\IEEEauthorblockA{\IEEEauthorrefmark{1}
Télécom SudParis - Institut Polytechnique de Paris, first-name.last-name@telecom-sudparis.eu}
\IEEEauthorblockA{\IEEEauthorrefmark{2}
Dept. of Electrical, Electronics and Information Engineering, University of Catania}
}

\maketitle

\begin{abstract}
Cloud Computing (CC) is the most prevalent paradigm under which services are provided over the Internet. The most relevant feature for its success is its capability to promptly scale service based on user demand. When scaling, the main objective is to maximize as much as possible service performance. Moreover, resources in the Cloud are usually so abundant, that they can be assumed infinite from the service point of view: an application provider can have as many servers it wills, as long it pays for it.

This model has some limitations. First, energy efficiency is not among the first criteria for scaling decisions, which has raised concerns about the environmental effects of today's "wild" computations in the Cloud. Moreover, it is not viable for Edge Computing (EC), a paradigm in which computational resources are distributed up to the very edge of the network, i.e., co-located with base stations or access points. In edge nodes, resources are limited, which imposes different \emph{parsimonious} scaling strategies to be adopted. 

In this work, we design a scaling strategy aimed to instantiate, \emph{parsimoniously}, a number of microservices sufficient to guarantee a certain Quality of Service (QoS) target. We implement such a strategy in a Kubernetes/Docker environment. The strategy is based on a simple Proportional-Integrative-Derivative (PID) controller. In this paper we describe the system design and a preliminary performance evaluation.
\end{abstract}

\section{Introduction}
\label{sec:intro}

Despite the many improvements in resource optimization for cloud computing, these remote systems fall victim to unexpected variations in service demands. In order to always guarantee service requirements, overprovising of computation resources can be adopted in the Cloud, with consequent economic and environmental costs. Due to the limited resources in the Egde, overprovisioning is not an option for Edge Computing (EC), where instead allocation strategies should be \emph{parsimonious}, i.e., use only the resources that are needed by a certain service and not more than those.


Microservice architectures have been proposed to easily maintain and scale a service with respect to the demand and have also been considered for EC~\cite{ma2017efficient}. While infrastructure owners/administrators  usually aim to guarantee a good QoS we believe that, additionally, microservice architectures are an opportunity to achieve \emph{parsimony} and finely tune the resource consumption of an application by dynamically allocating the right amount of microservices.
In our vision, parsimony is particularly important in edge nodes, where computational resources are limited (not infinite, as assumed in the cloud). In such cases, if multiple applications co-exists in the same edge node, it is crucial that each of them consumes the right amount of resources, not more, in order to leave room to the other applications.
At the same time, energy consumption can be dramatically reduced by reducing the number $P$ of processing microservices running~\cite{Hindle2018,7432984}. However, if the number $P$ of microservices is reduced too much, service requirements could not be met anymore. Therefore, a cautious control law must be adopted in order to find the ``right'' $P$. The calculation of this law must be light enough not to impact resource consumption at the edge.
The process of scaling up and down the number $P$ of microservices is called \emph{horizontal scaling}.

Our contribution in this paper is to demonstrate the feasibility of \emph{Parsimonious Edge Computing} (PEC). To this aim, we present a proof of concept of  a microservice architecture managed to limit the consumed resources. We implement it in a  Docker / Kubernetes-orchestrated environment, where clients generate a load of requests and multiple microservices can serve such requests. Requests are placed in a queue. A control law, based on Proportional-Integrative-Derivative (PID), performs horizontal scaling with the goal of maintaining such a queue on a certain target level. Differently from adopting PID in the Cloud, where we can afford complex monitoring systems~\cite{CerqueiraDeAbranches2016}, we propose a lightweight architecture.

Preliminary results obtained in a small scale testbed show that the system is able to dynamically orchestrate creation/removal of microservices in order to center the target.
Our architecture and experiments are reproducible and the code is released as open source~\cite{Simon2021}.

\section{Model}
\label{sec:methodology}




\begin{figure}[htp]
    \centering
    \includegraphics[width=8cm]{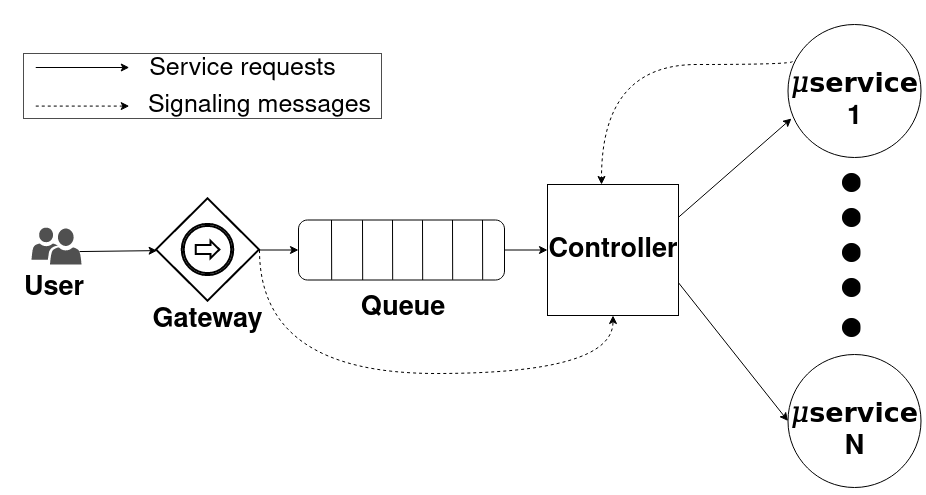}
    \caption{Design of the model}
    \label{fig:design}
\end{figure}



Fig.~\ref{fig:design} depicts the implemented system.
Users generate a sequence of service requests, e.g., antivirus calculations, or livecast video transcoding offloaded from end devices to an edge node~\cite{Offload1,Offloading2,JiangchuanLiu2017}. Requests are placed at the edge in a queue. There are $P$ microservices, denoted with $\mu$services hereafter, able to process such requests. While in our experiments they are Docker containers, they could also be ephemeral execution environments in the context of serverless computing. Immediately after a $\mu$service becomes free, it dequeues a request and serves it, which takes a certain processing time. In other words, our system is a multiserver queue. 

A \emph{Controller} dynamically adjusts the number $P$ of $\mu$services. The controller monitors at several instants $t$ the queue size $W(t)$, i.e., how many requests are still waiting in the queue to be served. Based on this observation, it adjusts the number $P$ of $\mu$services. This adjustment is calculated with a simple PID control law. In particular, the Proportional–Integral–Derivative controller (PID) issues a \emph{correction} $P_{out}(t)$ (see~\eqref{eq:Pout}), i.e., how many $\mu$services should be added (if $P_{out}>0$) or removed (if $P_{out}<0$). This correction $P_{out}(t)$ is added to the number $P(t)$ of currently deployed $\mu$services and gives the number of $\mu$services wanted (see~\eqref{eq:serice-wanted}), i.e., the ones that are necessary to drive the queue length $W(t)$ close to a pre-fixed target $T$ (which is equivalent to keep the error~\eqref{eq:error} as low as possible). 
The control law is summarized by the following equations:

\begin{small}
\begin{flalign}
\label{eq:serice-wanted}
 \mu\text{services wanted }  P_w(t) & = P(t) + P_{out}(t)
 \\
 \label{eq:Pout}
 \text{Control variable } P_{out}(t) 
 & = K_pe(t) + K_i\int_0^t{e(\tau)d\tau} + K_d\frac{d}{d\tau}e(t)
 \\
 \text{Error value } e(t) &= W(t) - T
 \label{eq:error}
\end{flalign}
\end{small}

By imposing the number $W(t)$ to be close to the target, we can limit the response time since we prevent the queue to explode. Consider a low latency application, with real time requirements, i.e. requests have deadlines of few milliseconds and failing to serve them within the deadline is equivalent to losing the requests. In this case, we can set the target queue size $T$ to a very small number. Our control law will aim to achieve such a target by deploying \emph{just the right amount}  of $\mu$services, in order not to over-use computational resources.

\section{Implementation}
We create a Docker/Kubernetes cluster that implements the concept of Parsimonious Edge Computing. In what follows, we denote with \emph{italic} font the \emph{service requests} generated by users (simply \emph{request} hereafter) and with \underline{underelined} font the \underline{signaling messages} needed to manage the system.

\paragraph{Service request handling.}
We emulate \textbf{users}, by generating a sequence of \emph{service requests} directed to the \textbf{Gateway}. Although we do not deal with encryption in this preliminary work, in the future extensions the Gateway would act as TLS termination proxy. This latter (i)~places the \emph{requests} into a Kafka \textbf{queue}~\cite{shree2017kafka} of one topic (therefore, the gateway acts as a ``Kafka producer'') and (ii)~sends a \underline{notification} to the Controller, via another HTTP GET, with an ID of the request. The gateway increments these IDs at every request so that consecutive requests have consecutive IDs. The Controller is a Docker container, which implements a web server able to handle HTTP requests. The controller is the only Kafka ``consumer'', i.e., it dequeues \emph{requests} from the queue via Kafka Python API~\cite{KafkaPython}. The Controller computes the queue length $W(t)$ as the difference between the ID of the last arrived \emph{request} in the queue and the ID of the lastly dequeued \emph{request}. The $\mu$service that processes the \emph{service requests} is a microservice implemented as a Docker container. We implement a pull-based mechanism: (i)~whenever the $\mu$service becomes free (it has finished to process the previous \emph{request}), it asks for a new \emph{requests} by sending an \underline{HTTP GET} to the Controller, (ii)~the Controller dequeues a \emph{request} from the queue and (iii)~sends \emph{it} to the $\mu$service via an \underline{HTTP Reply}.

\paragraph{Dynamic microservice deployment.}
Each $\mu$service has an ID assigned by the Controller, which keeps a list of active $\mu$services. 

Whenever the Controller receives a \underline{notification} by the Gateway of a new \emph{request} arriving in the queue, it checks the number $P(t)$ of active $\mu$services and compare it with the number $P_w(t)$ of $\mu$services wanted, established by PID control law. If $P_w(t)>P(t)$, the Controller creates $P_w(t)-P(t)$ new $\mu$services, assigning new IDs using Kubernetes primitives. Suppose that $t_1$ and $t_2$ are the arrival times of two consecutive requests. In the interval $t\in[t_1,t_2[$, by construction~\eqref{eq:serice-wanted}-\eqref{eq:error}, $W(t)\le W(t_1)$ because requests can only leave the queue. Therefore, $e(t)\le e(t_1)$, i.e., the PID law would never trigger creation of new $\mu$services between two service requests arrivals.

At any instant $t$ when a $\mu$service $s$ asks the Controller for a new \emph{service request}, the latter first checks if $P_w(t)<P(t)$: in this case, there are in the system more $\mu$services than needed and the controller destroys $\mu$service $s$ via Kubernetes primitives.  This mechanism ensures that we never destroy any $\mu$service that is currently processing a \emph{request}.

\section{Evaluation}
\label{sec:eval}

\setlength{\belowcaptionskip}{-0.6cm}

\begin{figure}
    \centering
    \includegraphics[width=8cm]{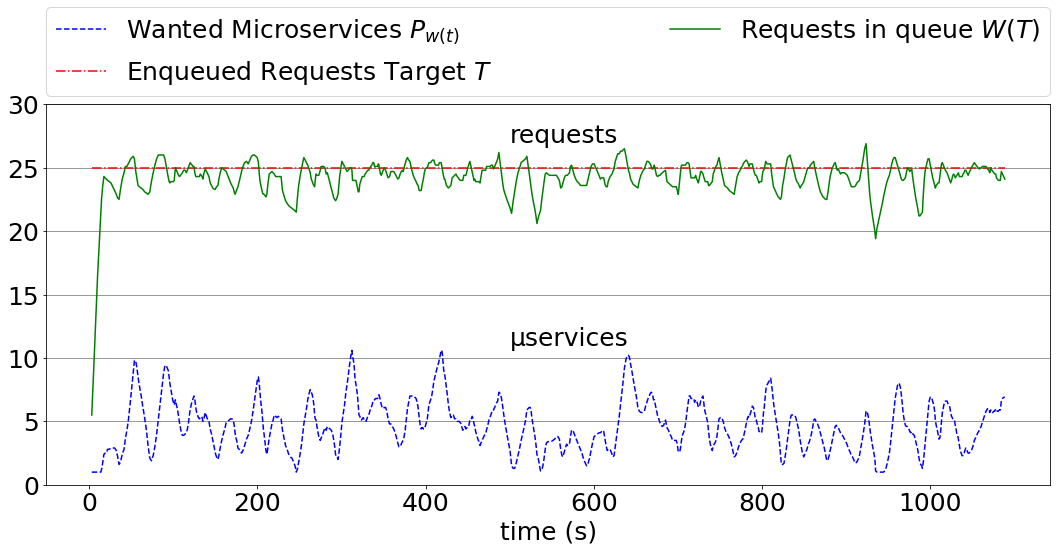}
    \caption{Number of requests in queue and $\mu$services (moving average, 10 samples per window.}
    \label{fig:results}
\end{figure}


We generate requests with exponentially distributed inter-arrival time with mean 1 second. The time for one $\mu$service to process one request is exponentially distributed with mean 5 seconds. We fix the target $T$ of requests waiting in the queue to 25. Based on observation, we tune the PID parameters to $K_p = -0.9$, $K_i = 0$ and $K_d = -0.2$.
Note that the creation of $\mu$services requires some non-negligible startup time~\cite{spoiala2016performance}, and thus the effects of the control values coming from the PID controller are delayed. Moreover, to keep you monitoring environment lightweight, we only get the current queue length $W(t)$ when new requests arrive or a $\mu$service asks for a new request to consume. Therefore, we sample the queue at non-regular time intervals. Moreover, there is an imbalance between creation and destruction of $\mu$services: we destroy them, at most one at a time, when a $\mu$service has finished processing its current request. On the other hand, we can create many $\mu$services in one shot. This results in the visible imbalance between the above-target and the below-target values of $W(t)$ in Fig.~\ref{fig:results}. Despite such difficulties, our system is able to get the queue length $W(t)$ around the target.

\section{Other horizontal scaling strategies for microservice architectures}
The subject of our work is horizontal autoscaling (Sec.~\ref{sec:intro}) to which we limit our discussion. Vertical autoscaling~\cite{rattihalli2019exploring} is out of scope.

The standard horizontal scaling strategy in Kubernetes~\cite{nguyen2020horizontal} requires the administrator to decide a threshold on CPU utilization. Once a $\mu$service exceeds such a threshold, new $\mu$services are created. However, there is no clear relation between application QoS and CPU utilization, so that it is difficult, if not impossible, for the administrator to set the right threshold for different classes of applications. As a consequences, if the administrator wants to be sure a certain QoS is satisfied, he needs to set stringent threshold, which may deploy more $\mu$services than actually needed, and which goes against the parsimony that we advocate here.

In Knative autoscaling~\cite{Knative2020}, instead, the administrator sets a ``concurrency'' value $x$, and the strategy deploys the amount of $\mu$services so that each one gets $x$ requests. As for the standard Kubernetes horizontal scaling, understanding what is the right value to set is challenging.  Similarly, OpenFaaS triggers $\mu$services based on thresholds, defined on incoming requests per second (RPS). It has been found that such concurrency- or RPS- based approaches are inefficient to meet request demand while limiting the number of triggered $\mu$services~\cite[Sec.~5]{li2019understanding}.

The horizontal scaling of a recent strategy, called Libra~\cite{Balla2020}, scales $\mu$services based on heuristic rules, which consists in thresholds on the requests served per second and request serving time. It will be interesting in our future work to compare our control-theoretical low with Libra's heuristic.

\section{Conclusion}
\label{sec:conclusion}
We have proposed a lightweigt architecture and control law for \emph{Parsimonious Edge Computing}.
We have shown an implementation based on Docker/Kubernetes/Kafka and released it as open source. We have numerically verified that our system manages to meet some pre-defined system-level performance target while limiting the amount of microservices, via dynamic allocation/destruction. In future work, we will compare our strategy with other horizontal scaling techniques (Sec.~5). We will implement with this model different classes of real services, with different requirements (low latency, high throughput). We will also consider different concurrent services running at the same node and competing for resources. Other control strategies, e.g., Reinforcement Learning, will be considered.




\section{Acknowledgment}
This work has been carried out in the context of the Chaire ``Les Réseaux du Future pour les Services de Demain'' at the Très Haut Débit (THD) Platform of Télécom SudParis.

\bibliographystyle{./bibliography/IEEEtran}
\bibliography{./bibliography/IEEEabrv,./bibliography/IEEEexample}

\end{document}